\documentclass{aastex}
\usepackage{spr-astr-addons}
\usepackage{url}

\RequirePackage{color}
\shorttitle{TAUVEX status: 2010 }
\shortauthors{Noah Brosch and Jayant Murthy}
\begin{document}


\title{TAUVEX: status in 2011}


\author{Noah Brosch}
\affil{The Wise Observatory and the Beverly and Raymond Sackler School of Physics and
Astronomy, Faculty of Exact Sciences, Tel Aviv University, Tel Aviv 69978, Israel}

\author{Jayant Murthy}
\affil{Indian Institute of Astrophysics, II Block Koramangala,
Bangalore 560 034, India}




%

\begin{abstract}

We present  a short history of the TAUVEX instrument, conceived to provide multi-band wide-field imaging in the ultraviolet, emphasizing the lack of sufficient and aggressive support on the part of the different space agencies that dealt with this basic science mission. First conceived in 1985 and selected by the Israel Space Agency in 1989 as its first priority payload, TAUVEX is fast becoming one of the longest-living space project of space astronomy. After being denied a launch on a national Israeli satellite, and then not flying on the Spectrum X-Gamma (SRG) international observatory, it was manifested since 2003 as part of ISRO's GSAT-4 Indian satellite to be launched in the late 2000s. However, two months before the launch,  in February 2010, it was dismounted from its agreed-upon platform. This proved to be beneficial, since GSAT-4 and its launcher were lost on April 15 2010 due to the failure of the carrier rocket's 3rd stage. TAUVEX is now stored in ISRO's clean room in Bangalore with no firm indications when or on what platform it might be launched.

\end{abstract}



\section{Introduction}\label{intro}
The Israel Space Agency (ISA) issued a call for pre-proposals in 1988  for ``scientific experiments to be flown on an Israeli satellite''. The call was answered by approximately 50 pre-proposals ranging from space astronomy experiments, to characterizing the behavior of electronic devices in the space environment, to the behavior of fishes in zero-g. All the pre-proposals were evaluated by an internal ISA panel and two were selected and funded to submit Phase A proposals. Among these, the one from Tel Aviv University proposed to orbit two small telescopes to provide relatively wide-field imaging in the space-ultraviolet (UV) domain.

The subsequent submission stage at the completion of the Phase A, performed together with a commercial contractor (El-Op Electro-Optical Industries now part of the ELBIT Systems company), produced a detailed study of the mission. El-Op was selected as Prime Contractor since it had a strong heritage of sophisticated electro-optical payloads for ground, naval, and airborne imaging while developing substantial infrastructure for space imaging payloads. In particular, it operated a thermal-vacuum chamber equipped with a collimator that allowed a payload to be end-to-end tested in vacuum and at extreme temperatures.

The TAUVEX Phase A result was a design summarized in Table 1, with three 20-cm co-aligned telescopes mounted within a cylinder that could fit the inner space of an OFEQ-class satellite. The field of view (FOV) of each telescope was chosen to be approximately one degree. Erring on the conservative side, we decided to use only space-proven techniques and components, and to require the Prime Contractor to include fully-redundant systems in this first national astronomy experiment. For UV detectors we selected sealed photoelectric detectors with Caesium Telluride semi-transparent cathodes equipped with multi-channel plate (MCP) intensifiers and wedge-and-strip anodes, as used in previous experiments (e.g., FAUST: Bowyer et al. 1993). At this time we decided to name the experiment TAUVEX, for {\bf T}el {\bf A}viv University {\bf UV} {\bf EX}periment. The science to be performed by TAUVEX, as proposed to ISA, was to survey a large fraction of the sky in the UV to depths vastly superior to those achieved by the UV surveys in the 1970s.

 The science goals of the TAUVEX mission were defined as providing galaxy imaging in a number of bands to allow studies of star formation processes, allowing the detection and identification of AGNs by selections based on color indices, and collecting unique stellar photometry data for stellar populations and interstellar extinction studies. These goals were discussed by e.g., Brosch \& Almoznino (2007), Joshi et al. (2007), Netzer (2007), Shastri (2007), and Maheswar et al. (2007). They were derived from the projected capabilities of TAUVEX given the properties listed by Brosch (1998) and subsequent modifications. Of special importance for the study of galaxies in the nearby Universe and of the stellar and interstellar matter of the Milky Way we mention the inclusion of two filters centered on the 2174\AA\, feature; photometric measurements through a wide and a narrow band filter allow the measurement of the equivalent width of this feature.

\begin{table}
\title{}

\end{table}

\begin{table*}
\caption{TAUVEX details}
\label{t:TAUVEX}
\begin{tabular}{ll} \hline
Property &	Description \\
      \hline
Number of telescopes & 3 \\
Aperture (individual telescope) & 20-cm \\
Optical design & Ritchey-Chr\'{e}tien \\
Operating spectral range & 120-320 nm \\
Filter wheels & four-position \\
Detector & Wedge-and-strip (CsTe$_2$; Peak QE$\simeq$10\%); photon-counting \\
Field of view & 56 arcmin \\
Electronic pixel size & 3 arcsec \\
Angular resolution & 6-11 arcsec (wavelength dependent) \\
Calculated sensitivity (SRG) & $\sim$20 mag (monochromatic) in 1$^h$ integration \\
Calculated sensitivity (GSAT-4) & $\sim$19 mag (monochromatic) while scanning the NPC \\
\hline
\end{tabular}

Note:  NPC=North Polar Cap: 90$^{\circ} \leq \delta \leq 80^{\circ}$ \\
\end{table*}

The TAUVEX Phase A report was submitted to the scrutiny of two evaluation committees: one technical, composed of Israeli technology experts to evaluate its feasibility, and the other scientific, composed of well-known international experts in the UV astronomy field to evaluate its scientific contribution and expected impact. The second selected proposal, submitted by the Technion, Israel's technological university, was evaluated at the same time; it proposed orbiting a small X-ray telescope.

The two evaluations proceeded in parallel and the conclusion was that ISA selected TAUVEX as the payload with the highest priority to launch. At this point, when the development, construction and launch of TAUVEX were expected to follow smoothly, ISA found that the national satellite and launcher on which it was counting to mount TAUVEX became unavailable. The reason, which became clear only in 2009, was that ISA originally planned to launch TAUVEX on the qualification model (QM) of one of the OFEQ-series satellite, the first three-axis-stabilized Israeli satellites used for imaging intelligence. The QM would have been refurbished and used to launch and operate TAUVEX which, as already described, was originally designed to fit the inner volume of the satellite.

However, with the loss at launch of one of the OFEQ satellites, practical necessities dictated the use of the satellite and launcher to orbit a different payload. With the disappearance of the satellite and launcher intended for pure scientific research, ISA announced that it could only provide funding for the development of the scientific instrument and that the science team should look for a platform on which TAUVEX could be flown.

\section{An alternative launch: SRG}\label{SRG}
With the help of international colleagues, the Spectrum X-$\gamma$ (SRG) spacecraft was identified as a possible carrier platform for TAUVEX. SRG was designed as a large high-energy astrophysics platform, one of the three major observatories in the Spectrum series, together with Spectrum RadioAstron and Spectrum-UV. The spacecraft was to be provided by the Soviet Union and constructed at the Lavotchkin Space Industries in Khimky (near Moscow), with most of its scientific instruments on-board to be supplied by different European countries. Originally, SRG was to be launched by a Proton rocket into a highly elliptical, four-day Molnyia-type orbit.

The largest instrument on-board SRG was the Danish-led SODART X-ray telescope with two 60-cm nested-cones concentrators for the $\sim$0.3-10 keV range (Schnopper 1990). SODART was to provide X-ray imaging over a $\sim$one degree FOV with arcmin angular resolution and excellent sensitivity, given its large-throughput optics. SRG included also soft X-ray to extreme UV imagers from Switzerland, another X-ray imager from the UK, a hard X-ray imager from Italy, etc.

The inclusion of TAUVEX among the SRG instruments was intended to provide simultaneous UV information with which the X-ray observations could be correlated, much in the manner of XMM's Optical/UV monitor. An additional task imposed on TAUVEX by the mission leaders was to aid the SRG fine guidance system in stabilizing the telescopes' LOS. This was to be done by providing fine guiding corrections to the SRG systems every two seconds; these corrections were derived from centroiding a relatively bright star in the TAUVEX FOV and measuring displacements from an initial position. Calculations showed that with the expected sensitivity TAUVEX could provide such stabilization signals all over the sky if it would have been used with no filter, or with a very wide blue-cutoff filter dubbed ``broad-band filter'' (BBF). Since this ``guiding'' task was deemed to be vital for SRG, the satellite Principal Investigator required TAUVEX to be equipped with a BBF filter in every telescope of the three it carried, albeit its scientific value was not expected to be significant.

The configuration of three telescopes within  a cylinder, conceived for an OFEQ launch, could not be maintained with TAUVEX mounted on top of SODART at the space end of SRG. For this reason, the planned cylindrical configuration was ``unwrapped'' and the three telescopes were designed to be almost coplanar. The payload was designed to fit into two units: an optical unit (OU) consisting of the mirrors, detectors, filter assemblies, and pre-amplifier boards, and an electronic unit (EU) consisting of the power conditioning unit, analog-to-digital front ends, high-voltage power supplies, doubly-redundant CPUs, on-board memory in the form of four 84 MB laptop hard disks in pressurized enclosures, etc. This configuration was described by Topaz et al. (1993), Schnopper (1994), Brosch et al. (1994), and Leibowitz (1995).

ISA and Israel's minister of Science and Technology (at that time the late Prof. Yuval Ne'eman) issued by the end of 1991 a letter to Prof. Rashid Sunyaev the SRG Principal Investigator promising to deliver TAUVEX to the SRG spacecraft on time, according to the satellite schedule. At this time, the plan was to launch SRG by 1994. With the commitment to launch on SRG, a contract was signed between ISA and the Ministry of Science and Technology on the one hand, and El-Op on the other hand, to provide the TAUVEX payload according to previously agreed-upon specifications. The investigator team at Tel Aviv University was to provide an advisory role in the project, with the final decisions to be taken by ISA.

Given the short time table, the design of TAUVEX and its construction were performed in record time by El-Op, and the project never missed any SRG milestone. This included the delivery of size and mass models and of a thermal model (TM), identical in shape to the flight model (FM) but equipped with many sensors so that it could fully simulate the behavior of TAUVEX in space conditions. Prior to its delivery to Lavotchkin, the TAUVEX TM undertook a full space simulation at the iABG facility in Germany, where a thermal-vacuum chamber equipped with a solar simulator was used. Note also that the pressure to launch in time required the assembly of the flight OU to take place already in 1993, with detectors fabricated in 1992.

However, the dissolution of the Soviet Union brought about financial difficulties in Russia which inherited most of the Soviet space programs. In particular, SRG suffered continuous delays and caused additional costs to the TAUVEX project. Parenthetically, note that SRG has not yet been launched by 2010, although it is still manifested as a Russian Space Agency (RKA) mission albeit with a different set of instruments. Given the interminable delays, and the uncertainty that SRG would ever launch, ISA instructed the TAUVEX science team in 2000 to search for an alternate launch possibility. A launch alternative was possible since TAUVEX was designed from the outset to require only mechanical fixation points, electrical power supply and uplink/downlink telemetry from any platform it would be attached to.

\section{Looking for an alternative launch: GSAT-4}\label{GSAT4}
The Indian Space Research Organization (ISRO) issued in 2000 a call for proposals to perform scientific experiments on its future GSAT-4 satellite. GSAT-4 was intended to be a technological demonstrator satellite for the next generation Indian telecommunications satellites testing, in particular, the ``bent-pipe'' transponder techniques and carrying a navigational payload. With apparently a large mass margin and intended to go into geosynchronous orbit (GO), GSAT-4 appeared to be a good carrier candidate for TAUVEX.

Discussions with the ISRO officials, and with the scientists at the Indian Institute of Astrophysics (IIA), resulted in identification of some necessary modifications to the SRG version of TAUVEX. In particular, since a telecom satellite in GEO must maintain a fixed orientation with respect to the Earth to keep its beams on the intended ground station, this required mounting TAUVEX on a rotating plate (MDP) able to aim the TAUVEX LOS to any declination on the celestial sphere. The ISRO engineers decided that they could realize this by mounting the MDP on a solar panel motor while routing the electrical lines through the motor spindle.

The advantage of having TAUVEX on a GEO platform is, obviously, the reduction of radiation pickup while going through the South Atlantic Anomaly or the inner radiation belts. This would have happened every four-day orbit in the first year of SRG operation. Another advantage of operation on-board a GEO telecom satellite is the availability, in principle, of continuous high-volume communications. The ISRO engineers realized this by modulating the GSAT-4 beacon; TAUVEX could thus enjoy continuous scientific telemetry at 1 Mbps without accessing any of the on-board experimental transponders.

After identifying the developments required for TAUVEX, a Memorandum of Understanding (MOU) was signed between ISRO and ISA on December 25 2003. The MOU stipulated specifically that TAUVEX will be launched on GSAT-4 and the signatories were the then heads of agencies: Sri Madhavan Nair for ISRO and Mr. Aby Har Even for ISA. The ceremony took place in Bangalore and was witnessed by Israel's Minister of Science and Technology (at that time, Mr. Eliezer Sandberg).

Following the MOU signing, TAUVEX was re-evaluated in 2005 by ISA and El-Op as to its readiness for flight. Since the FM OU had been assembled already almost a decade earlier, there was no way to test all its individual components. Instead, we relied on checking the response of the detectors to light and measuring the shape of the spectral response curves against the same curves provided by the detector manufacturer (DEP=Delft Electronische Producten in Roden, the Netherlands). We checked the filters separately by extracting the filter wheels from the OM. The transmission profiles checked out almost exactly as measured by the manufacturer and the detector outputs followed the spectral shapes produced by DEP more than a decade earlier. Similar results were obtained for the detector dark noise and hot pixels, concluding that both detectors and filters were flyable at their nominal performances.

We could not check the reflectivity of the mirrors since this would have required an end-to-end test, planned for the final testing and calibration, immediately prior to shipping TAUVEX to ISRO. Originally, El-Op was required to install witness pieces of the mirrors and contamination-monitoring germanium disks within TAUVEX; these were, however, not installed at any time after the FM's optical unit was assembled, precluding the possibility of checking for possible contamination or mirror coating degradation. The cleanliness of the FM OU was checked by MMG Sorek specialists, who performed the original environmental conditions evaluations (Noter et al. 1993, Nahor et al. 1993, Lifshitz et al. 1994). This was done by obtaining dry or wet ``smears'' from the stray light shield (inside and outside), from the mirror bezel, and from other external parts of TAUVEX. These revealed only minor contamination levels and, as a conclusion, predicted that the UV performance in space would be the nominal one.

\section{Modifications for GSAT-4}\label{GSAT4mods}
Changing from a stabilized pointing mode to one in which the telescopes rotate about the Earth necessitated changes in the data handling system. In the SRG configuration, the hard disk memories had only to record images (with 1000$\times$1000 pixel resolution, one for each telescope) for each observation period. In the new geosynchronous orbit, each photon event had to be recorded with its coordinates on the detector plane and exact time. Knowing the pointing angle at that time, an image could later be reconstructed from all the recorded events. With a launch scheduled for 2004 or later, it made sense also to renew many of the elements in the FM EU so as to enhance the payload reliability. For instance, the four 84-MB hard disks used as on-board storage for the SRG version were replaced by 2$\times$4 GB solid-state memories to provide a buffer storage for instances when the photon event rate would exceed the capabilities of the telemetry link. On the hardware side, since the installation on the MDP allowed a slight extension of the OU, an external baffle was designed to fit in front of the old stray light shield. This 10-cm extension was calculated to reduce further the influence of extraneous light and reduce the background, at the expense of a slight mass increase.

Major changes took place in the heat dissipation system; this now would be done by dedicated radiators viewing deep space from next to the apertures of the additional baffle. The heat was conducted by heat pipes embedded in the MDP via copper ribbons from the OU detector compartments, and by direct contact with the bottom of the EU.

The flight software was extensively rewritten, eliminating the significant on-board data processing required for a flight on SRG. In particular, the image reconstruction required to reduce the telemetry bandwith on SRG was no longer needed; instead, it was necessary to download each photon event, with time tags inserted in the data stream every 0.128 sec.

\section{Possible problems:  uncertain ground testing and calibration}\label{qual}
In preparation for the first Flight Model integration session with GSAT-4 (following an integration exercise performed with the TAUVEX Engineering Model), scheduled for the second half of 2008 for a launch scheduled for early-2009, we performed the acceptance tests and the final ground calibrations of the TAUVEX FM at the El-Op facilities in spring 2008. The procedures and the derived results were described in detail by Almoznino et al. (2009).

Briefly, we found that while TAUVEX complied with most of the requirements as defined in the ``System Specifications'' document, which is part of the contract between ISA and El-Op, there were a number of properties where the results deviated significantly from the desired performance.

The most disturbing finding was that the overall throughput seemed to be only 10-20\% of the nominal one and is demonstrated by the two plots in Fig. 1. The nominal throughput was calculated assuming two mirror reflections at 85\%, five transmissive surfaces at 98\%, filter transmissions as measured, and the typical cathode quantum efficiency following DEP. This reduced throughput result was, however, highly uncertain because the testing equipment at El-Op was not UV-optimized. For instance, the thermal vacuum chamber collimator was coated for the visible spectral range and its mirror coatings had not been checked for reflectivity for more than a decade. Because of this and of the numerous intermediate calibration stages required to perform the entire process, we estimated the uncertainties in the throughput reduction factor to be close to 100\% for some of the far-UV bands (see also Almoznino et al. 2009). Such large uncertainties allowed, in principle, also for nominal performance, though the more likely results were lower than nominal.

\begin{figure}[htbp]
\label{RealResponse_pix}
 \includegraphics[width=65mm,angle=-90]{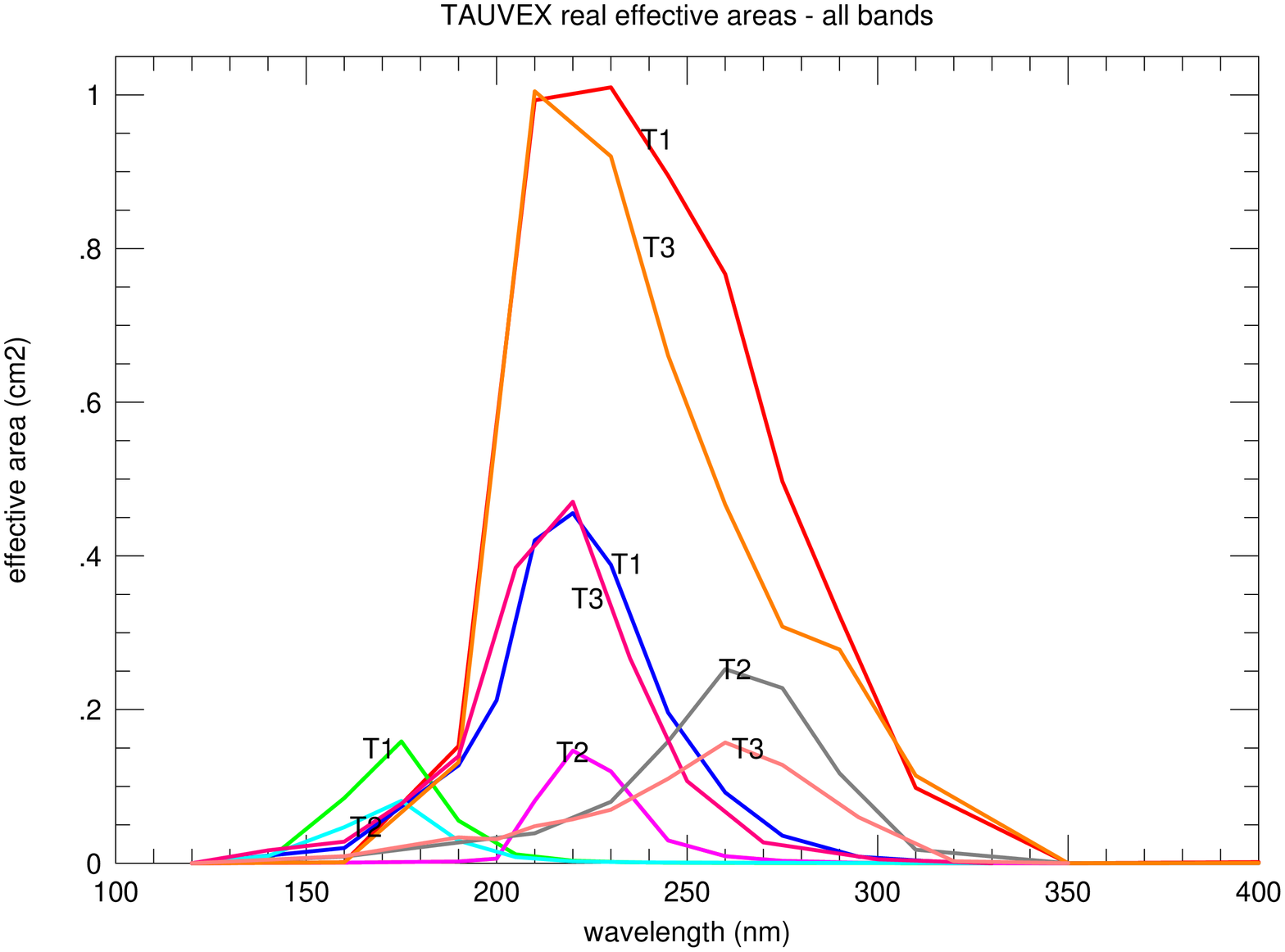}
  \includegraphics[width=65mm,angle=0]{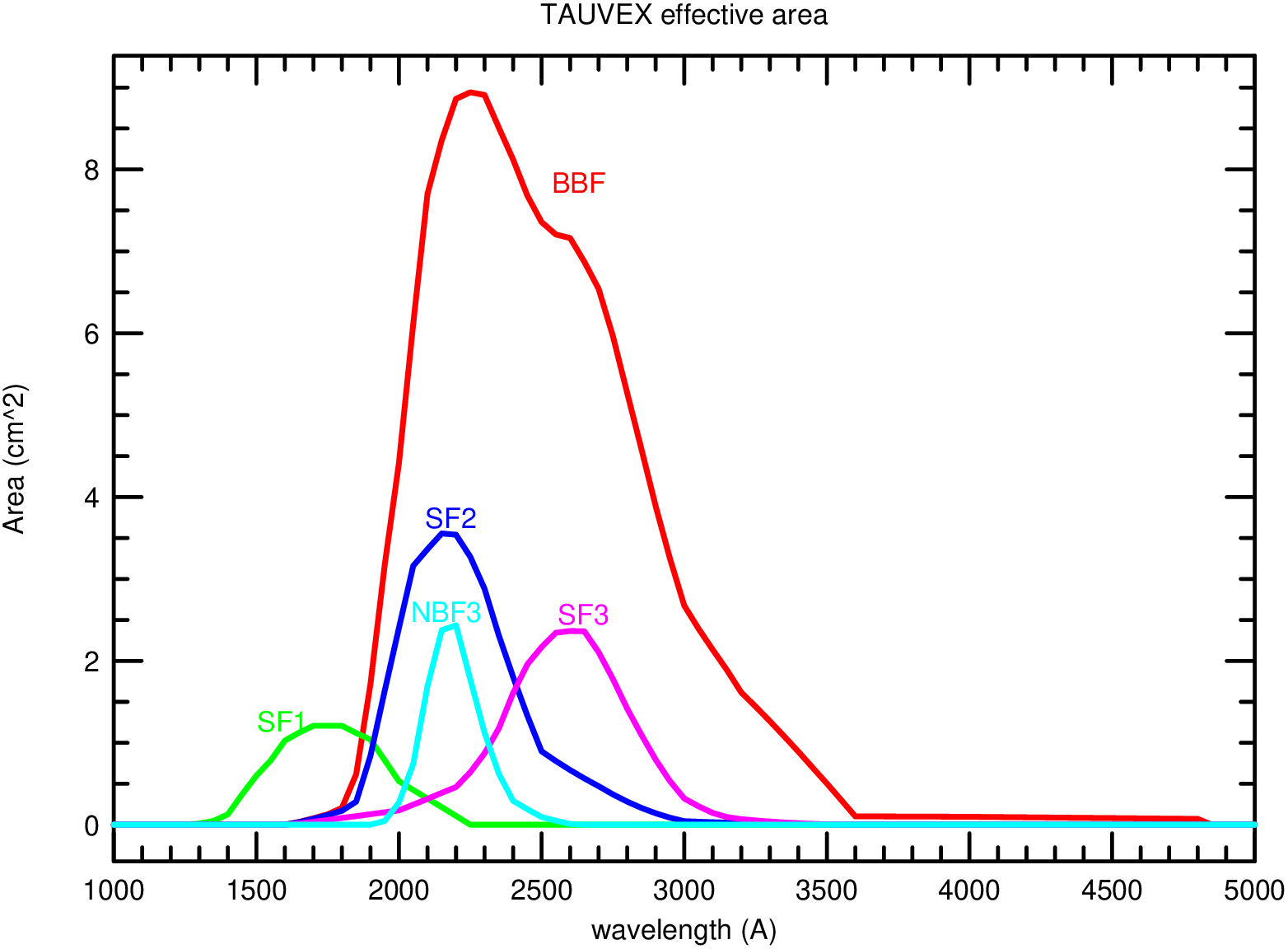}
  \caption{{\it Upper panel:} Measured response of the TAUVEX FM telescopes in the different bands, expressed as ``effective area'' in cm$^2$. The filters color-coded and are labeled on the lower plot. {\it Lower panel:} Expected response of the generic TAUVEX telescopes in the different bands, calculated on the basis of individual component characteristics.}
\end{figure}

The conclusion that the throughput was probably lower than expected was strengthened by the findings of the stray light rejection calibration. Although performed only at 20, 40 and 50$^{\circ}$ away from the LOS whereas TAUVEX was designed to operate with the Sun at least 90$^{\circ}$ from the LOS, the values obtained during the calibration were almost two orders of magnitude lower than those theoretically calculated. Since miracles hardly happen, we concluded that we might indeed have a case of lowered overall sensitivity. As the acceptance criterion for overall throughput called for at least 90\% of the nominal, the implication was that TAUVEX failed its acceptance test.

The explanation offered by the El-Op technical team concerning the loss in system sensitivity pointed to the telescope mirrors as likely culprits. These were coated by evaporated films of aluminum and magnesium fluoride to protect the aluminum. It is possible that the humidity at which these mirrors were kept for 15 years, $\sim$50\% in the clean room of El-Op, caused the deterioration of the MgF$_2$ protective film and this allowed the diffusion of air contaminants, perhaps oxygen molecules or other compounds, to the aluminum layer that lost part of its reflecting capability. This explanation is supported by the findings of Fern$\acute{\rm a}$ndez-Perea et al. (2006) who showed that the degradation of an Al+MgF$_2$ combination exposed to N$_2$, O$_2$, or H$_2$O was mostly due to the water molecules; these affect also the transmittance of the MgF$_2$ protective layer.

While the original TAUVEX sensitivity was designed and expected to yield photometry accurate to 0.1 mag or better for sources brighter than $\sim$20 mag (Vega system) in the UV for an integration of $\sim$200 sec, that is, a single pass through the detector along its diameter, the new results indicated this level of accuracy could be achieved only for UV sources brighter than 17-18 mag. The possible remedy, of recoating the mirrors then re-assembling TAUVEX and re-calibrating, was deemed not to be feasible given the high costs and the schedule, which called for a 2009 launch.

The Indian and Israeli science teams resolved to compensate for the possible loss of sensitivity by accumulating longer on-source integrations, by scanning preferentially regions near the two celestial poles while reducing the sky area to be covered. At high declinations the dwell time of a source in the FOV increases as 1/cos$\delta$ with the telescopes scanning at diurnal rate, reaching a full day of integration at the pole itself. We therefore devised a program for the first two months of performance verification when we would carefully calibrate TAUVEX as to sensitivity, stray light, etc., essentially performing an acceptance test using calibrated celestial sources. Following this period, if the throughput would be found as low as during the ground tests, we planned to spend most of the time on the North and South celestial polar caps ($|\delta|\geq80^{\circ}$), performing a survey of these areas to the depth of the GALEX MIS with the suspected reduced sensitivity, or to the level of the GALEX DIS if the original sensitivity would be confirmed in-orbit.

\section{Integrations with GSAT-4}\label{GSAT4integration}
Various delays at ISRO moved the first integration exercise of TAUVEX to the end of 2008. The integration took place at the ISRO clean room facilities in Bangalore with the different TAUVEX flight units on a table and the satellite units mounted in the satellite but connected to TAUVEX via cables that simulated the on-board telemetry. This allowed a first testing of the command and scientific telemetry and of the electrical connections to the spacecraft. For this test the spacecraft power was supplied from UPS devices; at the second integration the supply was from solar panel simulators and undesired TAUVEX behaviour was detected as will be described below.

TAUVEX was first tested alone against the test equipment (EGSE) brought with it from Israel. It was then connected to the satellite subsystems with the satellite open and long cables connecting TAUVEX to the beacon modulator (for downlink scientific telemetry at 1 Mbps) and to the various other connectors for uplink command and downlink slow (technical) telemetry. In general, this exercise showed that the design was correct and that most of the systems operated as planned; those that did not could be reworked promptly to do so for the subsequent integration stage.

The second and final pre-flight integration, originally expected to take place in early-2009 for a launch before summer 2009, was delayed to November 2009. This integration was with the complete satellite, although some units not related to TAUVEX were still to be mounted. The procedure required less than four days and could be completed in such a short time due to the dedication of the El-Op team and the extensive and expert help tended by the ISRO specialists. The latter helped significantly in debugging the various problems that appeared during the integration.



A full end-to-end test of TAUVEX and GSAT-4 was performed at the completion of the integration. This included moving the MDP to preset angles and measuring the LOS, acquiring diffuse light images and sending them to the EGSE through the GSAT-4 telemetry system, etc. All these functioned perfectly, except that whenever GSAT-4 was operated from the solar panel simulator instead of a UPS strange patterns appeared in the TAUVEX images and the point spread functions of unresolved pinhole sources became excessively large. It was clear that we had a noise pickup problem in the TAUVEX power supply; this was fixed by adding a filtering stage to the TAUVEX power input at the GSAT-4 end.

It is worth stressing that at the end of the final integration not only was TAUVEX itself ready for launch, but also the science teams were ready for the mission in software, mission planning and science planning. In particular, exercises of the two teams were held to prove the reduction pipeline, including the reconstruction of the UV images collected by the three telescopes using artificial event streams as well as real data collected during the ground calibrations. A detailed mission plan for the first flight months was also generated; it included the space acceptance tests discussed above in \S~\ref{qual}.

\section{Problems with GSAT-4}
Since spring 2009 rumors began to filter to the TAUVEX team about possible problems with the GSAT-4 mission. These were never more than word-of-mouth statements having to do with uncertainties regarding the capabilities of the launcher. Nevertheless, at the completion of the final pre-launch integration stage, performed to the satisfaction of both ISRO and El-Op teams, the El-Op engineers and the TAUVEX scientists were convinced that only a few more steps had to be completed in the near future before starting to enjoy the scientific output from TAUVEX: a final vibration and acoustic noise test of the complete satellite including TAUVEX, and a partial test of TAUVEX with GSAT-4 at the launch site, prior to the satellite integration with the GSLV vehicle. The launch was supposed to follow promptly in January or February 2010.

In December 2009, ISRO officials contacted the TAUVEX team and suggested that it might be advisable to take TAUVEX off the satellite because of concerns that the satellite mass was too high to allow a lifetime longer than 6 months in orbit. In addition, there were suggestions based on an earlier GSAT flight that light scattered by the GSAT-4 solar panels would adversely affect observations toward the celestial poles, where the best TAUVEX data would have come from. The TAUVEX team was not able to independently quantify this additional contribution to the scattered light.

During these discussions with the ISRO officials, they mentioned a possibility to launch TAUVEX on a small dedicated satellite that could be lofted to low Earth orbit with a PSLV launch. This option, that appeared to the science teams to be genuine, would essentially have restored TAUVEX to its original mission as proposed to ISA and described in \S~\ref{intro}.

Although neither ISA nor Tel Aviv University or El-Op ever agreed to the removal of TAUVEX from GSAT-4, by late-January or early-February 2010 the MDP with TAUVEX were unloaded from GSAT-4 and the satellite was closed up for launch. TAUVEX was left in the ISRO clean room, protected by clean anti-static plastic sheets.

\section{Launch and failure}
ISRO decided to launch GSLV carrying GSAT-4 on 15 April 2010 despite the possibility that the cryogenic upper stage of the launcher would not function properly. The launch was broadcast live and the Israeli team viewed it in real time. About 500 seconds following ignition, after the first and second stages completed their burning, the third (cryogenic) upper stage lit up but apparently could not sustain the burn. The rocket was seen not accelerating, taking a nose dip, losing attitude control, and presumably crashing into the ocean.

The TAUVEX team experienced mixed feelings. On the one hand they grieved for the loss of the satellite and launcher, and for the waste of so many years of hard work by the dedicated ISRO teams. On the other hand, they felt an obvious sense of relief that TAUVEX was spared, remaining safe in the ISRO clean room, and expecting to fly on a future platform. However, up to the time this paper was written, the inter-agency contacts did not yet yield a date or a specific platform that could loft TAUVEX to perform its intended scientific tasks. In the meantime, TAUVEX was dismounted from the MDP by an El-Op team and was securely stored in its transportation container that provides an optimal storage environment since it is filled with dry nitrogen. The container with TAUVEX is stored in the ISRO clean room in Bangalore.

Given the delay till a possible launch would become available, it became clear to the science teams of both nations that any flight requiring a delay longer than one year would have to be undertaken with a TAUVEX performing as originally planned. Since the mirrors had been singled as the likely source of degradation, they would have to be recoated. This, fortunately, could be done in India and would not require reshipping the payload to Israel, then back to India.

\section{Discussion} \label{Discussion}
The prolonged TAUVEX saga described above begs a number of conclusions regarding the structure of the TAUVEX project managed in Israel. The first deals with omissions by the Israeli side; it was a mistake on the part of the science team to agree to continue the project after the national satellite and launcher promised by ISA in the original call for proposals became unavailable. This naive approach, that provided no exit strategies throughout the duration of the project, is understandable on the part of scientists, but ultimately did not pay off.

The second mistake on the part of the TAUVEX Principal Investigators was to allow ISA to take full control of the budget and of the final decisions regarding the conduct of the project at El-Op, while relegating themselves to an advisory role. During the long years of this project this caused the relaxation of the tasks the Prime Contractor (El-Op) was required to perform, such as the installation of contamination monitoring devices within TAUVEX. Similarly, not insisting on a thorough investigation of the implications of the long-term storage of the completed TAUVEX OU was a mistake and the improper storage caused the throughput reduction presumably by degrading the mirror coatings. ISA also relaxed the requirement to maintain full documentation for the project; the conclusion is that now there is no way to know what was done with the flight model, when, or by whom.

On the Indian side, ISRO carries significant blame since it entered into an agreement with ISA to launch TAUVEX on-board GSAT-4 but was not able to fulfil it. ISA could not enforce the launch agreements with either RKA or ISRO because no punitive consequences were included in the launch agreements. ISRO misled, willingly or unwillingly, both science teams, Indian and Israeli, as to the status of the GSLV launcher and of the GSAT-4 satellite. The delays certainly did not help the state of the TAUVEX mirrors and caused unnecessary and significant expenses to the Israeli side. If the one-year delay (2008-2009) in the integration would have been known in advance, it is possible that El-Op would have had time to consider refurbishing the TAUVEX optics to recover the original response instead of keeping TAUVEX in the clean room of ISRO.

  However, one should also mention here the successes of the project, in particular (a) the built-in flexibility of the payload that allowed a relatively easy shift from SRG to GSAT-4, and (b) the strict adherence of the Prime Contractor (El-Op) to the different schedules imposed by the two satellites. Additionally, the cooperation of the technical teams at both Lavotchkin Industries in Russia and ISITE in India facilitated the testing at various stages and the easy and efficient integration of the FM with GSAT-4.

With an eye to the future, the Indian and Israeli science teams expect ISA and ISRO to reach an agreement very soon to allow a fully-recovered TAUVEX to be launched and perform its original mission in a timely manner. In particular, the science team proposed to refurbish and recoat the mirrors; this is expected to recover the original sensitivity of TAUVEX. However, when this paper was completed no such decision, or indeed {\bf any} decision, was announced by either ISA or ISRO.

In the second decade of the 21st century, when the GALEX mission is drawing to a conclusion with its FUV channel not functioning since 2009 and UVIT on ASTROSAT not yet launched, the science community is in dire need of a wide-field high-sensitivity UV imaging mission. The UV imagers on SWIFT and XMM provide only a limited capability in comparison to that of a recovered TAUVEX. HST has exquisite UV imaging capabilities but these are limited to very small fields. 

While not useful for understanding the high-redshift Universe (because of the redshift) or studying exoplanets (angular resolution, visit cadence), TAUVEX could easily fulfill an important role as a wide-angle UV imager for the world astronomical community and bring into play some of its special features that, so far, have not been duplicated by other UV experiments. One such example is the use of two of its filters, NBF-3 and SF-2, to measure the equivalent width of the 2174\AA\, feature of interstellar dust (Brosch 1996). Other examples, e.g. in the field of galaxy evolution (Brosch \& Almoznino 2007) or AGN research (Netzer 2007), abound. In addition, given that its collecting areas per telescope are smaller that GALEX, TAUVEX could observe brighter sources than GALEX without endangering the detectors or going into non-linear detector response. Such sources have been avoided by the GALEX mission and are now ``holes'' in the UV sky (e.g., the Magellanic Clouds).

\section{Acknowledgements}
We are grateful to the many people who read and commented on various drafts of this paper. Valuable remarks from an anonymous referee significantly improved the paper. All the involved individuals and agencies were given the opportunity to remark on the draft; some did and we are grateful for their contributions.

\end{document}